\documentclass[prd,onecolumn,superscriptaddress,showpacs,nofootinbib,preprintnumbers]{revtex4}

\usepackage{amsmath}
\usepackage{amsfonts}
\usepackage{graphicx}
\usepackage{dcolumn}
\usepackage{hyperref}

\def\tx{\tilde{x}}

\def\be{\begin{equation}}
\def\ee{\end{equation}}
\def\bea{\begin{eqnarray}}
\def\eea{\end{eqnarray}}

\def\d{{\rm d}}

\def\5{\overline 5}
\def\vp{\varphi}

\begin{document}

\title{Coupled dark energy: Towards a general description of the 
dynamics}

\author{Burin Gumjudpai}
\affiliation{Fundamental Physics \& Cosmology Research Unit,
The Tah Poe Group of Theoretical Physics (TPTP), \\
Department of Physics, Naresuan University, Phitsanulok,
Thailand 65000}
\author{Tapan Naskar}
\affiliation{IUCAA, Post Bag 4, Ganeshkhind, Pune 411 007, India}
\author{M.~Sami}
\affiliation{IUCAA, Post Bag 4, Ganeshkhind,
Pune 411 007, India}
\author{Shinji Tsujikawa}
\affiliation{Department of Physics, Gunma National College of
Technology, Gunma 371-8530, Japan}

\date{\today}

\vskip 1pc
\begin{abstract}

In dark energy models of scalar-field coupled to 
a barotropic perfect fluid, the existence of cosmological scaling solutions restricts
the Lagrangian of the field $\vp$ to 
$p=X g(Xe^{\lambda \vp})$, where $X=-g^{\mu\nu} \partial_\mu \vp
\partial_\nu \vp /2$, $\lambda$ is a constant and $g$ is an arbitrary function.
We derive general evolution equations in an autonomous form 
for this Lagrangian and investigate the stability of fixed points
for several different dark energy models--(i) ordinary (phantom) field,
(ii) dilatonic ghost condensate, and (iii) (phantom) tachyon.
We find the existence of scalar-field dominant fixed 
points ($\Omega_\vp=1$) with an accelerated expansion
in all models irrespective of the presence of the 
coupling $Q$ between dark energy and dark matter.
These fixed points are always classically stable for a phantom field,
implying that the universe is eventually dominated by the energy 
density of a scalar field if phantom is responsible for dark energy.
When the equation of state $w_\vp$ for the field $\vp$ is larger than $-1$, 
we find that scaling solutions are stable if the scalar-field dominant 
solution is unstable, and vice versa. 
Therefore in this case the final attractor is either a scaling solution 
with constant $\Omega_\vp$ satisfying $0<\Omega_\vp<1$ or a scalar-field 
dominant solution with $\Omega_\vp=1$. 

\end{abstract}

\pacs{98.70.Vc}

\maketitle
\vskip 1pc

\section{Introduction}

Over the past few years, there have been enormous efforts in constructing models of 
dark energy--either motivated by particle physics or 
by phenomenological considerations 
(see Refs.~\cite{review} for review). 
The simplest explanation of dark energy is provided by cosmological constant, 
but the scenario is plagued by a severe fine tuning problem associated with its energy scale.
This problem can be alleviated by considering a scalar field
with a dynamically varying equation of state.
In the recent years, a host of scalar field dark energy models have been proposed, 
ranging from quintessence \cite{Paul}, k-essence \cite{Kes},  
Born-Infeld scalars \cite{Tachyonindustry},
phantoms \cite{phantom0,phantom}, ghost condensates \cite{Arkani,PT} etc. 

In a viable dark energy scenario we require that the 
energy density of scalar field remains subdominant during radiation and 
matter dominant eras and that it becomes important only at late times 
to account for the current acceleration of universe.
In this sense cosmological {\it scaling solutions} can be
important building blocks in constructing the models of dark 
energy \cite{CLW,LS,scaling,TS}. 
The energy density of a scalar field $\vp$ decreases proportionally 
to that of a barotropic perfect fluid for scaling solutions.
Steep exponential potentials give rise to scaling solutions
for a minimally coupled scalar field in General Relativity allowing
the dark energy density to mimic
the background fluid during radiation or matter dominant 
era \cite{BCN}. If the field potential $V(\vp)$ becomes less steep
at some moment of time, the universe exits from the scaling regime 
and enters the era of an accelerated expansion \cite{BCN,Sahni}.

The quantity, $\lambda \equiv -V_{\vp}/V$, which characterizes the slope of the potential,
is constant for exponential potentials.
In this case it is straightforward to investigate the stability of critical points in 
phase plane \cite{CLW}. Even for general potentials a similar 
phase space analysis can be done by considering 
``instantaneous'' critical points with a dynamically changing 
$\lambda$ \cite{scaling}. Therefore we can understand 
the basic structure of the dynamics of dark energy by 
studying the fixed points corresponding to scaling solutions.
Note that the potentials yielding scaling solutions are different depending upon 
the theories we adopt. For example we have 
$V(\vp) \propto \vp^{-2/(n-1)}$ \cite{TS}
for the universe characterised by a Friedmann equation: $H^2 \propto \rho^n$
[the Randall-Sundrum (RS) braneworld and the RS Gauss-Bonnet 
braneworld correspond to $n=2$ and $n=2/3$, respectively].
The scaling solution for tachyon corresponds to the inverse square potential:
$V(\vp) \propto \vp^{-2}$ \cite{Abramo,AL,CGST}.

Typically dark energy models are based on scalar fields minimally coupled to gravity and 
do not implement the explicit coupling of the field to a background fluid. 
However there is no fundamental reason for this assumption
in the absence of an underlying symmetry which would suppress the coupling. 
The  possibility of a scalar field $\vp$ coupled to a matter and its cosmological consequences 
were originally pointed out in Refs.~\cite{Ellis}.
Amendola proposed a quintessence scenario coupled with dark matter 
\cite{Luca} as an extension of nonminimal coupling theories \cite{Luca2}
(see Ref.~\cite{Doran} for an explicit coupling of a quintessence 
field to fermions or dark matter bosons).
It is remarkable that the scaling solutions in coupled 
quintessence models can lead to a late-time acceleration, while this is not 
possible in the absence of the coupling.
Recently there have been attempts to study the dynamics of a phantom 
field coupled to dark matter \cite{Guo,NOT}. 

In Refs.\,\cite{PT,TS} it was shown that the existence of
scaling solutions for coupled dark energy 
restricts the form of the field Lagrangian to 
be $p(X, \vp)=X\,g(Xe^{\lambda \vp})$,
where $X=-g^{\mu \nu}\partial_\mu \vp \partial_\nu
\vp/2$ and $g$ is any function in terms of $Xe^{\lambda \vp}$.
This result is very general, since it was derived by starting from a
general Lagrangian $p(X, \vp)$ which is an arbitrary
function of $X$ and $\vp$. 
In fact this Lagrangian includes a wide variety of dark energy models such as
quintessence, phantoms, dilatonic ghost condensates and
Born-Infeld scalars.
While the critical points corresponding to scaling solutions were derived in 
Ref.\,\cite{PT,TS}, this is not sufficient to understand the 
properties of all the fixed points in such models.
In fact scaling solutions correspond to the fractional energy density of scalar fields
satisfying $0<\Omega_\vp<1$, but it is known that $\Omega_\vp$
approaches 1 for several fixed points in the cases of
an ordinary field \cite{CLW} and 
a tachyon field \cite{Abramo,AL,CGST} when the coupling $Q$ is absent 
between dark energy and dark matter.

Our aim in this paper is to study the fixed points and their stabilities against perturbations
for coupled dark energy models with the field Lagrangian: $p(X, \vp)=X\,g(Xe^{\lambda \vp})$.
This includes quintessence, dilatonic ghost condensate and tachyons,
with potentials corresponding to scaling solutions.
We shall also study the case of phantoms with a negative kinematic term
in order to understand the difference from normal scalar fields. 
While phantoms are plagued by the problem of vacuum instability at the 
quantum level \cite{UV}, we would like to clarify the classical stability around 
critical points. We note that this quantum instability for phantoms is overcome
in the dilatonic ghost condensate scenario provided that higher-order derivative 
terms stabilize the vacuum \cite{PT}.

The rest of the paper is organised as follows. In Sec.\ II we briefly review the formalism
of a general scalar field $\vp$  coupled to barotropic fluid
and establish the autonomous form  of evolution equations for the Lagrangian 
$p(X, \vp)=X\,g(Xe^{\lambda \vp})$.
In Sec.\ III, we apply our autonomous equations to the system with 
a standard (phantom) scalar field, obtain all the critical points and 
investigate their stabilities.
Sec.\ IV and Sec.\ V are devoted to the detailed phase space analysis of dilatonic ghost condensate
and (phantom) tachyon field, respectively.
In Sec.\ VI we bring out some generic new features of coupled dark energy scenarios.

\section{Scalar-field model}

Let us consider scalar-field models of 
dark energy with an energy density $\rho$
and a pressure density $p$. 
The equation of state for dark energy is defined by 
$w_\vp \equiv p/\rho$.
We shall study a general situation in which a field $\vp$ responsible 
for dark energy is coupled to a barotropic perfect fluid 
with an equation of state: $w_m \equiv p_m/\rho_m$.

In the flat Friedmann-Robertson-Walker background with a scale 
factor $a$, the equations for $\rho$ and $\rho_m$ are \cite{PT}
\bea
\label{rho}
& & \dot{\rho}+3H(1+w_\vp)\rho=-Q \rho_m \dot{\vp}\,, \\
& & \dot{\rho}_m+3H(1+w_m)\rho_m=+Q \rho_m \dot{\vp}\,,
\label{rhom}
\eea
where $H \equiv \dot{a}/a$ is the Hubble rate with a dot being 
a derivative in terms of cosmic time $t$.
The equation for the Hubble rate is 
\bea
\label{dotH}
\dot{H}=-\frac12 \left[(1+w_\vp)\rho+(1+w_m)\rho_m \right]\,,
\eea
together with the constraint
\bea
\label{Hubble}
3H^2=\rho+\rho_m\,.
\eea
Here we used the unit $8\pi G=1$ ($G$ is a gravitational constant).
In Eqs.~(\ref{rho}) and (\ref{rhom}) we introduced a coupling $Q$ between 
dark energy and barotropic fluid by assuming the interaction given in Ref.~\cite{Luca}.
In Refs.~\cite{Guo,NOT} the authors adopted different forms of the coupling.
In what follows we shall restrict our analysis to the case of 
positive constant $Q>0$, but  
it is straightforward to extend the analysis 
to the case of negative $Q$.

We define the fractional density of dark energy and barotropic fluid,
$\Omega_\vp \equiv \rho/(3H^2)$ and $\Omega_m \equiv \rho_m/(3H^2)$, 
with $\Omega_\vp+\Omega_m=1$ by Eq.~(\ref{Hubble}).
Scaling solutions are characterized by constant values of $w_\vp$
and $\Omega_\vp$ during the evolution.
Then the existence of scaling solutions restricts the form of the 
scalar-field pressure density to be \cite{PT,TS}
\bea
\label{lag}
p=X\,g(X e^{\lambda \vp})\,.
\eea
where $X=-g^{\mu\nu} \partial_\mu \vp
\partial_\nu \vp /2$ and $g$ is any function in terms of $Y \equiv 
Xe^{\lambda \vp}$. Here $\lambda$ is defined by 
\bea
\label{lQ}
\lambda \equiv Q \frac{1+w_m-\Omega_\vp (w_m-w_\vp)}
{\Omega_\vp (w_m-w_\vp)}\,,
\eea
which is constant when scaling solutions exist.
$\lambda$ is related with the slope of the scalar-field potential $V(\phi)$.
For example one has 
$\lambda \propto -V_\vp/V$ for an ordinary field \cite{scaling} and 
$\lambda \propto -V_\vp/V^{3/2}$ for a tachyon 
field \cite{CGST}.
Then the associated scalar-field potentials are given by 
$V=V_0e^{-\lambda \vp}$ for the ordinary field and 
an inverse power-law potential $V=V_0\phi^{-2}$ 
for the tachyon.

For the Lagrangian (\ref{lag}) the energy density for the field $\vp$ is 
\bea 
\rho=2X\frac{\partial p}{\partial X}-p
=X\left[g(Y)+2Yg'(Y)\right]\,, 
\eea
where a prime denotes the derivative in terms of $Y$.
Then Eq.~(\ref{rho}) can be rewritten as 
\bea 
& &\left[g(Y)+5Yg'(Y)+2Y^2g''(Y)\right]\ddot{\vp}
+3H\left[g(Y)+Yg'(Y)\right] \dot{\vp}
+\lambda XY \left[3g'(Y)+2Yg''(Y)\right]=-Q\rho_m\,.
\eea
We introduce the following dimensionless quantities:
\bea
\label{xandy}
x \equiv \frac{\dot{\vp}}{\sqrt{6}H}\,,~~~~
y \equiv \frac{e^{-\lambda \vp/2}}{\sqrt{3}H}\,.
\eea
{}From this definition, $y$ is positive (we do not consider 
the case of negative $H$). 

Defining the number of $e$-folds as $N \equiv {\rm ln}\,a$,
we can cast the evolution equations in the following autonomous form:
\bea 
\label{dx}
\frac{\d x}{\d N} &=&   
x\left[ \frac32 (1+w_m)+\frac32 (1-w_m)x^2 g(Y)-
3w_m x^2 Yg'(Y)-\frac{3\{g(Y)+Yg'(Y)\}}
{g(Y) + 5Yg'(Y) + 2Y^2 g''(Y)}\right]
\nonumber \\
& &-\sqrt{6}\frac{\lambda x^2 Y \{3g'(Y)+2Yg''(Y)\}+Q[1-x^2\{g(Y)+2Yg'(Y)\}]}
{2[g(Y) + 5Yg'(Y) + 2Y^2 g''(Y)]}\,,\\
\label{dy}
\frac{\d y}{\d N} &=&   
-\frac{\sqrt{6}}{2}\lambda xy+\frac32 y
\left[1+w_m+(1-w_m)x^2g(Y)-2w_m x^2 Yg'(Y)\right]\,, \\
\label{dHN}
\frac{1}{H}\frac{\d H}{\d N} &=&   
-\frac32 \left[1+w_m+(1-w_m)x^2g(Y)
-2w_m x^2 Yg'(Y)\right]\,.
\eea
We also find
\bea 
\label{Ome}
\Omega_\vp =x^2[g(Y)+2Yg'(Y)]\,,~~~~
w_\vp =\frac{g(Y)}{g(Y)+2Yg'(Y)}\,.
\eea
It is also convenient to define the total effective equation 
of state:
\bea 
\label{weff}
w_{\rm eff}\equiv \frac{p+p_m}{\rho+\rho_m}
=w_m+(1-w_m)x^2g(Y)-2w_m x^2 Yg'(Y)\,.
\eea
Combining Eq.~(\ref{dHN}) with Eq.~(\ref{weff})  we obtain
\bea 
\frac{\ddot{a}}{aH^2}=-\frac{1+3w_{\rm eff}}{2}\,.
\eea
This means that the universe exhibits an accelerated expansion 
for $w_{\rm eff}<-1/3$.

It may be noted that the  quantity $Y$ is constant along a scaling solution, i.e., $Y=Y_0$.
However $Y$ is not necessarily conserved for other fixed points
for the system given by Eqs.~(\ref{dx}) and (\ref{dy}).
Therefore one can not use the property $Y=Y_0$ in order to 
derive the fixed points except for scaling solutions.
In subsequent sections we shall apply the evolution equations 
(\ref{dx}) and (\ref{dy}) to several different dark energy models--
(i) ordinary (phantom) scalar field, (ii) dilatonic ghost condensate
and (iii) (phantom) tachyon.

By using Eqs.~(\ref{lQ}), (\ref{Ome}), and (\ref{weff}), one can 
show that $w_{\rm eff}$ is written as
\bea 
w_{\rm eff}=\frac{w_m\lambda -Q}{Q+\lambda}\,,
\eea
whose form is independent of the function $g(Y)$.
For the pressureless fluid ($w_m=0$) 
we have $w_{\rm eff}=-Q/(Q+\lambda)$.
Then we obtain $w_{\rm eff} \to -1$
in the limit $Q \gg \lambda$.

{}From Eq.~(\ref{Ome}) we find that the $Q$-dependent term in 
Eq.~(\ref{dx}) drops out when $\Omega_\vp$ approaches 1.
Therefore, if the fixed point corresponding to $\Omega_\vp=1$
exists for $Q=0$, the same fixed point should appear even 
in the presence of the coupling $Q$.
This is a general feature
of any scalar field system mentioned above and would necessarily
manifest in all models we consider in the following sections.

\section{Ordinary (phantom) scalar field}

It is known that a canonical 
scalar field with an exponential potential
\bea 
\label{normalp}
p(X, \vp)=\epsilon X-c e^{-\lambda \vp}\,,
\eea
possesses scaling solutions.
We note that $\epsilon=+1$ corresponds to a standard field and 
$\epsilon=-1$ to a phantom.
In fact this Lagrangian can be obtained by starting with 
a pressure density of the form: $p=f(X)-V(\vp)$ \cite{PT,TS}.
We obtain the Lagrangian (\ref{normalp}) by choosing 
\bea 
\label{nscalar}
g(Y)=\epsilon- c/Y\,,
\eea
in Eq.~(\ref{lag}).
In what follows we shall study the case with $\lambda>0$ without the loss of
generality, since negative $\lambda$
corresponds to the change $\vp \to -\vp$ in Eq.~(\ref{normalp}).

For the choice (\ref{nscalar}) Eqs.~(\ref{dx}) and (\ref{dy})
reduce to 
\bea 
\label{dxnor}
\frac{\d x}{\d N} &=&   
-3x+\frac{\sqrt{6}}{2} \epsilon \lambda c y^2+\frac32
x \left[(1-w_m)\epsilon x^2 +(1+w_m)(1-c y^2)\right]
-\frac{\sqrt{6}Q}{2}\epsilon (1-\epsilon x^2 -c y^2)\,, \\
\frac{\d y}{\d N} &=&   
-\frac{\sqrt{6}}{2}\lambda xy+\frac32 y
 \left[(1-w_m)\epsilon x^2 +(1+w_m)(1-c y^2)\right]\,,
 \label{dynor} 
\eea
where $Y$ is expressed through $x$ and $y$, as
$Y=x^2/y^2$.
We note that these coincide with those given in 
Ref.~\cite{CLW} for $\epsilon=+1$ and $Q=0$.
Eqs.~(\ref{Ome}) and (\ref{weff}) give 
\bea 
\Omega_\vp=\epsilon x^2+c y^2\,,~~~
w_\vp=\frac{\epsilon x^2-c y^2}
{\epsilon x^2+c y^2}\,,~~~
w_{\rm eff}=w_m+(1-w_m)\epsilon x^2-
(1+w_m)c y^2\,.
\eea
\subsection{Fixed points}

The fixed points can be obtained by setting $\d x/\d N=0$
and $\d y/\d N=0$ in Eqs.~(\ref{dxnor}) and (\ref{dynor}).
These are presented in Table I.

\begin{itemize}

\item (i) Ordinary field ($\epsilon=+1$)

The point (a) gives some fraction of the field energy density
for $Q \ne 0$. However this does not provide an accelerated expansion, 
since the effective equation of state $w_{\rm eff}$ is positive 
for $0 \le w_m<1$. 
The points (b1) and (b2) correspond to kinetic dominant 
solutions with $\Omega_\vp=1$ and do not satisfy the 
condition $w_{\rm eff}<-1/3$.
The point (c) is a scalar-field dominant solution 
($\Omega_\vp=1$), which gives an acceleration of the universe
for $\lambda^2<2$. The point (d) corresponds to a cosmological 
scaling solution, which satisfies $w_\vp=w_m$
for $Q=0$. When $Q \ne 0$
the accelerated expansion occurs for $Q>\lambda (1+3w_m)/2$.
We note that the points (b1), (b2) and (c) exist irrespective of 
the presence of the coupling $Q$, since $\Omega_\vp=1$
in these cases.

\item (ii) Phantom field ($\epsilon=-1$)

It is possible to have an accelerated expansion for the point (a)
if the condition, $Q^2>(1-w_m)(1+3w_m)/2$, is satisfied.
However this case corresponds to an unphysical situation, 
i.e., $\Omega_\vp<0$ for $0 \le w_m<1$.
The critical points (b1) and (b2) do not exist for the phantom field.
Since $w_{\rm eff}=-1-\lambda^2/3<-1$ for the point (c), the universe
accelerates independent of the values of $\lambda$ and $Q$.
The point (d) leads to an accelerated expansion for 
$Q>\lambda (1+3w_m)/2$, as is similar to the case of  
a normal field.

\end{itemize}

\subsection{Stability around fixed points}

We study the stability around the critical points
given in Table I.
Consider small perturbations $u$ and $v$ about the
points $(x_c, y_c)$, i.e.,
\begin{eqnarray}
x=x_c+u\,,~~~y=y_c+v\,.
\label{uv}
\end{eqnarray}
Substituting into Eqs.~(\ref{dx}) and (\ref{dy}),
leads to the first-order differential equations for 
linear perturbations: 
\begin{eqnarray}
\frac{\d}{\d N}  
\left(
\begin{array}{c}
u \\
v
\end{array}
\right) = {\cal M} \left(
\begin{array}{c}
u \\
v
\end{array}
\right) \ ,
\label{uvdif}
\end{eqnarray}
where ${\cal M}$ is a matrix that depends upon
$x_c$ and $y_c$. 
The elements of the matrix ${\cal M}$ for the model 
(\ref{normalp}) are 
\begin{eqnarray}
a_{11} &=& -3+\frac92 \epsilon x_c^2 (1-w_m) +
\frac32 (1+w_m)(1-cy_c^2)+\sqrt{6}Qx_c\,, \\
a_{12} &=& \sqrt{6}\epsilon c \lambda y_c-
3c (1+w_m)x_cy_c+\sqrt{6}\epsilon c Qy_c\,, \\
a_{21} &=& -\frac{\sqrt{6}}{2}\lambda y_c+
3\epsilon x_cy_c(1-w_m)\,, \\
a_{22} &=&  -\frac{\sqrt{6}}{2}\lambda x_c-
3c (1+w_m)y_c^2+\frac32 (1-w_m)\epsilon x_c^2
+\frac32 (1+w_m)(1-cy_c^2)\,.
\end{eqnarray}
One can study the stability around the fixed points by 
considering the eigenvalues of the matrix ${\cal M}$.
The eigenvalues are generally given by 
\bea
\label{eigen}
\mu_{1, 2}=\frac12 \left[a_{11}+a_{22} \pm
\sqrt{(a_{11}+a_{22})^2-4(a_{11}a_{22}-a_{12}a_{21})}
\right]\,.
\eea
We can evaluate $\mu_1$ and $\mu_2$ for each critical point:

\begin{itemize}

\item Point (a):  
\[
\mu_1=-\frac32 
(1-w_m)+\frac{Q^2}{\epsilon(1-w _m)}\,,~~~
\mu_2=\frac{1}{\epsilon (1-w_m)}
\left[ Q(\lambda+Q)+\frac{3\epsilon}{2}
(1-w_m^2) \right]\,.
\]

\item Point (b1): 
\[
\mu_1=3-\frac{\sqrt{6}}{2}\lambda\,,~~~
\mu_2=3(1-w_m)+\sqrt{6}Q\,.
\]

\item Point (b2): 
\[
\mu_1=3+\frac{\sqrt{6}}{2}\lambda\,,~~~
\mu_2=3(1-w_m)-\sqrt{6}Q\,.
\]

\item Point (c): 
\[
\mu_1=\frac12 (\epsilon \lambda^2-6)\,,~~~
\mu_2= \epsilon \lambda (\lambda+Q) -3(1+w_m)\,.
\]

\item Point (d): 
\[
\mu_{1, 2}=-\frac{3\{\lambda(1-w_m)+2Q\}}
{4(\lambda+Q)} \left[1 \pm \sqrt{1+\frac{8[3(1+w_m)-\epsilon
\lambda (\lambda+Q)][3\epsilon (1-w_m^2)+2Q(\lambda+Q)]}
{3\{\lambda (1-w_m)+2Q\}^2}} \right]\,.
\]

\end {itemize}

\begin{table*}[t]
\begin{center}
\begin{tabular}{|c|c|c|c|c|c|}
Name & $x$ & $y$ & $\Omega_\vp$ & $w_\vp$ & 
$w_{\rm eff}$  \\
\hline
\hline
(a) & $-\frac{\sqrt{6}Q}{3\epsilon (1-w_m)}$ & 0 & 
$\frac{2Q^2}{3\epsilon (1-w_m)}$ &  1 &
$w_m+\frac{2Q^2}{3\epsilon (1-w_m)}$  \\
\hline
(b1) & $\frac{1}{\sqrt{\epsilon}}$ & 0 & 1 & 1 &
1 \\
\hline
(b2) & $-\frac{1}{\sqrt{\epsilon}}$ & 0 & 1 & 1 & 
1 \\
\hline
(c) & $\frac{\epsilon \lambda}{\sqrt{6}}$ & 
$[\frac{1}{c}(1-\frac{\epsilon \lambda^2}{6})]^{1/2}$
 & 1 & $-1+\frac{\epsilon \lambda^2}{3}$ &
 $-1+\frac{\epsilon \lambda^2}{3}$ \\
\hline
(d) & $\frac{\sqrt{6}(1+w_m)}{2(\lambda+Q)}$ & 
$[\frac{2Q(\lambda+Q)+3\epsilon (1-w_m^2)}
{2c(\lambda+Q)^2}]^{1/2}$ & 
$\frac{Q(\lambda+Q)+3\epsilon (1+w_m)}{(\lambda+Q)^2}$ & 
$\frac{-Q(\lambda+Q)+3\epsilon w_m (1+w_m)}
{Q(\lambda+Q)+3\epsilon (1+w_m)}$ &
$\frac{w_m \lambda -Q}{\lambda +Q}$ \\
\hline
\end{tabular}
\end{center}
\caption[crit1]
{\label{crit1} The critical points for the ordinary 
(phantom) scalar field. The points (b1) and (b2) do not exist 
for the phantom field.}
\end{table*}

One can easily verify that the above eigenvalues coincide with 
those given in Ref.~\cite{CLW} for $\epsilon=+1$ and $Q=0$.
The stability around the fixed points can be generally 
classified in the following way: 

\begin{itemize}

\item (i) Stable node: $\mu_1<0$ and $\mu_2<0$.

\item (ii) Unstable node: $\mu_1>0$ and $\mu_2>0$.

\item (iii) Saddle point: $\mu_1<0$ and $\mu_2>0$ (or $\mu_1 >0$ and 
$\mu_2<0$).

\item (iv) Stable spiral: The determinant of the matrix ${\cal M}$, i.e., 
${\cal D} \equiv (a_{11}+a_{22})^2-4(a_{11}a_{22}-a_{12}a_{21})$,
is negative and the real parts of $\mu_1$ and $\mu_2$ are negative. 

\end {itemize}

In what follows we shall analyze the stability around the fixed points
for the ordinary field and the phantom.

\subsubsection{Ordinary field $(\epsilon=+1)$}

When the coupling $Q$ is absent, the system with an 
exponential potential was already investigated in Ref.~\cite{CLW}.
When $Q \ne 0$, the stability of fixed points was 
studied in Ref.~\cite{Luca} when dust and radiation are present.
We shall generally discuss the property of fixed points for a fluid 
with an equation of state: $0 \le w_m<1$.

\begin{itemize}

\item Point (a): 

In this case $\mu_1$ is negative
if $Q < \sqrt{3/2}(1-w_m)$ and positive otherwise. 
Meanwhile $\mu_2$ is positive for any value of $Q$ and 
$\lambda$ (we recall that we are considering the case of 
positive $Q$ and $\lambda$). Therefore this point 
is a saddle for $Q < \sqrt{3/2}(1-w_m)$
and an unstable node for $Q > \sqrt{3/2}(1-w_m)$.
We note that the condition $\Omega_\vp<1$ gives 
$Q<\sqrt{(3/2)(1-w_m)}$.
Therefore the point (a) is a saddle for $w_m=0$
under this condition.

\item Point (b1): 

While $\mu_2$ is always positive, $\mu_1$ is 
negative if $\lambda>\sqrt{6}$ and positive otherwise.
Then the point (b1) is a saddle for $\lambda>\sqrt{6}$ and 
an unstable node for $\lambda<\sqrt{6}$.

\item Point (b2): 

Since $\mu_1$ is always positive and $\mu_2$
is negative for $Q > (3/2)^{1/2}(1-w_m)$ and positive otherwise, 
the point (c) is either saddle or an
unstable node.

\item Point (c): 

The requirement of the existence of the point (c)
gives $\lambda<\sqrt{6}$, which means that $\mu_1$
is always negative. The eigenvalue $\mu_2$ is negative
for $\lambda<(\sqrt{Q^2+12(1+w_m)}-Q)/2$ and positive otherwise.
Therefore the point (c) presents a stable node for $\lambda 
<(\sqrt{Q^2+12(1+w_m)}-Q)/2$, whereas 
it is a saddle for $(\sqrt{Q^2+12(1+w_m)}-Q)/2<\lambda<\sqrt{6}$.

\item Point (d): 

We first find that $-3\{\lambda(1-w_m)+2Q\}/4(\lambda+Q)<0$ 
in the expression of $\mu_1$ and $\mu_2$.
Secondly we obtain $\lambda(\lambda+Q)>3(1+w_m)$
from the condition, $\Omega_\vp<1$.
Then the point (d) corresponds to a stable node for 
$3(1+w_m)/\lambda-\lambda<Q<Q_*$ and a stable spiral 
for $Q>Q_*$, where $Q_*$ satisfies the following relation
\begin{eqnarray}
\label{Qstar}
3\left[\lambda (1-w_m)+2Q_*\right]^2=
8\left[\lambda (\lambda+Q_*)-3(1+w_m)\right]
\left[2Q_*(\lambda+Q_*)+3(1-w_m^2) \right]\,.
\end{eqnarray}
For example we have $Q_*=0.868$
for $\lambda=1.5$ and $w_m=0$.

\end {itemize}

The stability around the fixed points and the condition for an 
acceleration are summarized in Table II.
The scaling solution (d) is always stable provided that $\Omega_\vp<1$, 
whereas the stability of the point (c) is dependent on the values 
of $\lambda$ and $Q$.
It is important to note that the eigenvalue $\mu_2$ for the 
point (c) is positive when the condition for the existence 
of the point (d) is satisfied, i.e., $\lambda(\lambda+Q)>3(1+w_m)$.
Therefore the point (c) is unstable for the parameter range of 
$Q$ and $\lambda$ in which the scaling solution (d) exists.

\begin{figure}
\includegraphics[height=3.0in,width=2.8in]{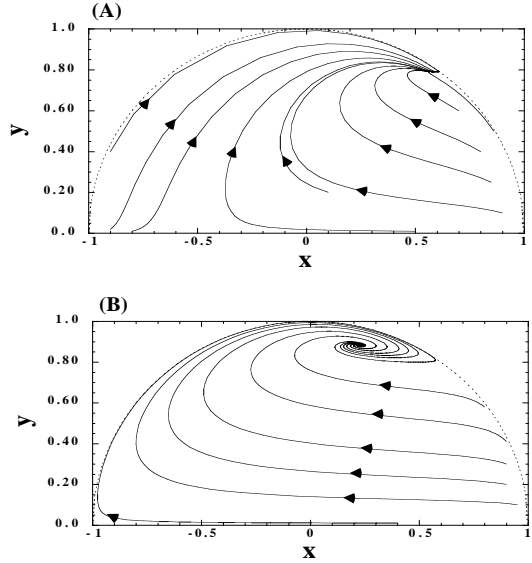}
\caption{\label{ordinary}
The phase plane for a standard scalar field corresponding to 
$Q=0.6$ [panel (A)] and $Q=4.0$ [panel (B)]  for $\lambda=1.5$, 
$w_m=0$ and $c=1$. The late-time attractor corresponds to a
stable node for $Q=0.6$ and a stable spiral for $Q=4.0$.
The dotted curve is $x^2+y^2=1$, which characterizes
the border of the allowed region ($\Omega_\vp=1$).
}
\end{figure}

In Fig.~\ref{ordinary} we plot the phase plane for $\lambda=1.5$, 
$w_m=0$ and $c=1$ with two different values of $Q$.
In the phase space the allowed range corresponds to $0 \le x^2+y^2 \le 1$.
When $\lambda=1.5$ the point (c) is a saddle for $Q>0.5$,
and the point (d) is a stable node for 
$0.5<Q<0.868$ and a stable spiral for $Q>0.868$.
The panel (A) in Fig.~\ref{ordinary} corresponds to the phase plane 
for $Q=0.6$, in which case the point (d) is a stable node. 
We find that all trajectories approach the stable node (d), i.e.,
$x_c=0.5832$ and $y_c=0.8825$.
In the panel (B) of Fig.~\ref{ordinary} we plot 
the phase plane for $Q=4.0$. It is clear that the critical point 
(d) [$x_c=0.2227$ and $y_c=0.8825$] is a stable spiral 
as estimated analytically.

\begin{table*}[t]
\begin{center}
\begin{tabular}{|c|c|c|c|}
Name & Stability & Acceleration & Existence \\
\hline
\hline
(a) & Saddle point for $Q < (3/2)^{1/2}(1-w_m)$  & No 
& $Q<(3/2)^{1/2}(1-w_m)^{1/2}$ 
\\
&  Unstable node for $Q > (3/2)^{1/2}(1-w_m)$ & &  \\
\hline
(b1) & Saddle point for $\lambda >\sqrt{6}$  & No &
All values  \\
&  Unstable node for $\lambda < \sqrt{6}$ & &\\
\hline
(b2) & Saddle point for $Q > (3/2)^{1/2}(1-w_m)$  & No & 
All values \\
&  Unstable node for $Q < (3/2)^{1/2} (1-w_m)$ & &\\
\hline
(c) & Saddle point for $([Q^2+12(1+w_m)]^{1/2}-Q)/2<\lambda<\sqrt{6}$  
& $\lambda<\sqrt{2}$ & $\lambda<\sqrt{6}$    \\
&  Stable node for $\lambda<([Q^2+12(1+w_m)]^{1/2}-Q)/2$ & &\\
\hline
(d) & Stable node for $3(1+w_m)/\lambda-\lambda<Q<Q_*$  
& $Q>\lambda (1+3w_m)/2$ & $Q>3(1+w_m)/\lambda-\lambda$  \\
&  Stable spiral for $Q>Q_*$ & &
\\
\hline
\end{tabular}
\end{center}
\caption[crit2]{\label{crit2} The conditions for stability \& acceleration
\& existence for an ordinary scalar field ($\epsilon=+1$).
We consider the situation with positive values of $Q$ and $\lambda$.
Here $Q_*$ is the solution of Eq.~(\ref{Qstar}).
}
\end{table*}

%
\subsubsection{Phantom field $(\epsilon=-1)$}
%

\begin{table*}[t]
\begin{center}
\begin{tabular}{|c|c|c|c|}
Name & Stability & Acceleration  & Existence \\
\hline
\hline
(a) & Saddle point for $Q(Q+\lambda) < (3/2)(1-w_m^2)$  
& $Q^2>(1-w_m)(1+3w_m)/2$ & No if the condition 
$0 \le \Omega_\vp \le 1$ 
\\
&  Stable node for $Q(Q+\lambda) > (3/2)(1-w_m^2)$ & &
is imposed
\\
\hline
(c) & Stable node & All values & All values  \\
\hline
(d) & Saddle & Acceleration for $Q>\lambda (1+3w_m)/2$ & 
$Q(Q+\lambda)>(3/2)(1-w_m^2)$ 
\\
\hline
\end{tabular}
\end{center}
\caption[crit3]{\label{crit3} 
The conditions for stability \& acceleration
\& existence for a phantom scalar field ($\epsilon=-1$).
We consider the situation with positive values of $Q$ and $\lambda$.
}
\end{table*}

\begin{figure}
\includegraphics[height=2.8in,width=3.0in]{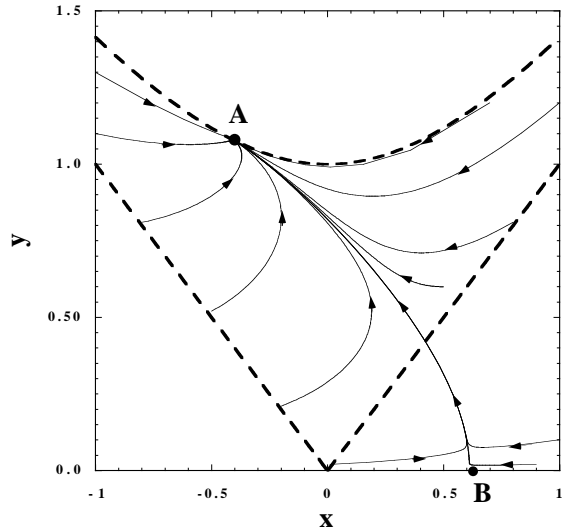}
\caption{\label{phan1}
Phase plane for a phantom-type scalar field ($\epsilon=-1$)
for $Q=3/4$, $\lambda=1$, $w_m=0$ and $c=1$.
The point A is a stable fixed point (c),
whereas the point B is the fixed point (a) corresponding to a saddle.
All trajectories approach the point A, which gives 
$w_{\rm eff}=-4/3$ and $\Omega_\vp=1$.
We also show the border of the allowed range ($-1 \le x^2-y^2 \le 0$)
as dotted curves.
}
\end{figure}

\begin{figure}
\includegraphics[height=2.8in,width=3.0in]{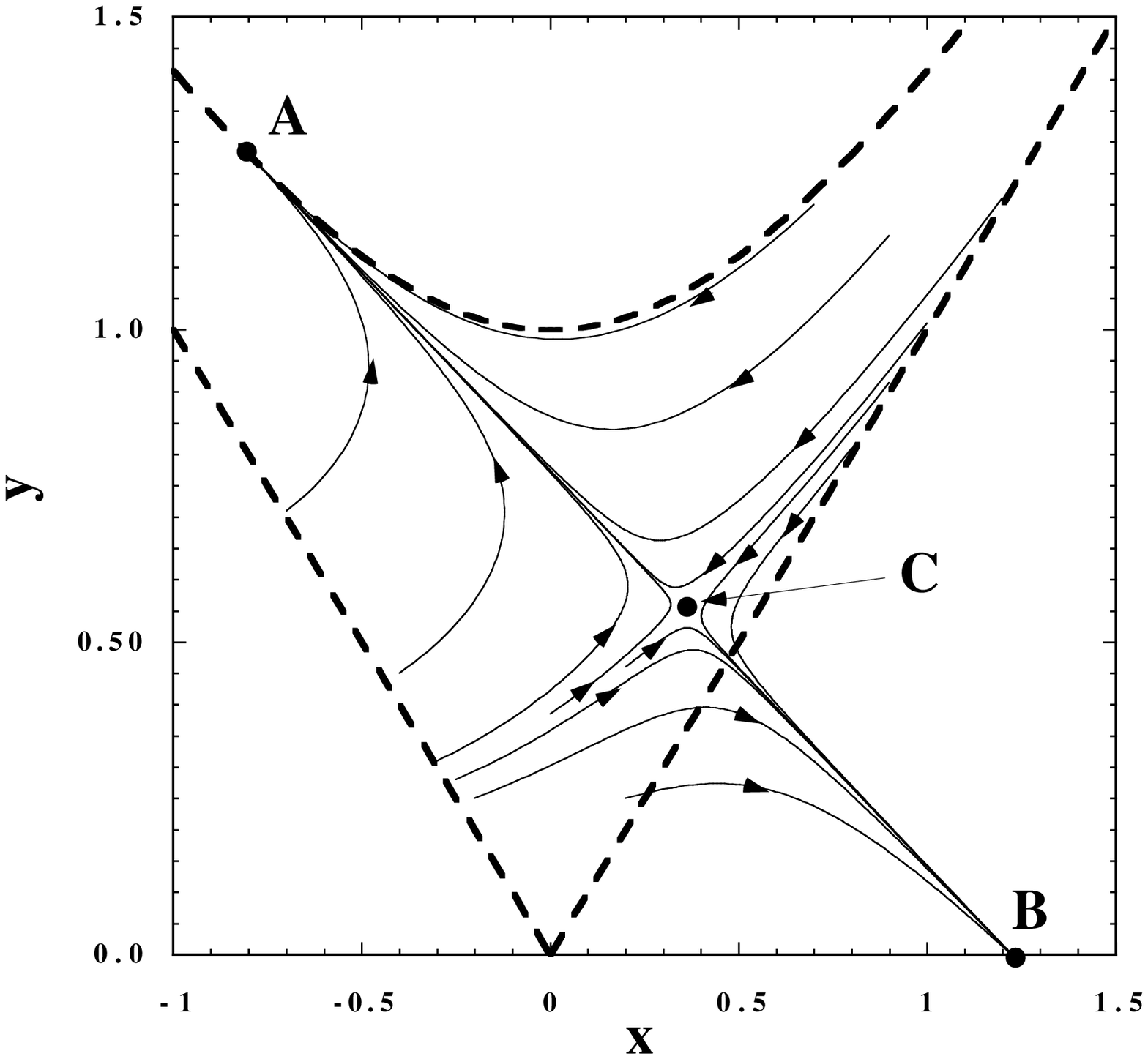}
\caption{\label{phan2}
Phase plane for a phantom-type scalar field ($\epsilon=-1$)
corresponding to $Q=3/2$, $\lambda=2$, $w_m=0$ and $c=1$.
The point A is a stable fixed point (c),
whereas the point B is the fixed point (a) corresponding to a stable node.
The point C is the fixed point (d) corresponding to a saddle.
In this case the trajectories approach either the point A or B, 
depending on the initial conditions of $x$ and $y$.
The border of the allowed range is the same as in Fig.~\ref{phan1}.
}
\end{figure}

The fixed points (b) and (c) do not exist for the phantom field.

\begin{itemize}

\item Point (a): 

In this case $ \mu _1$ is always negative, 
whereas $\mu _2$ can be either positive
or negative depending the values of $Q$ and $\lambda$.
Then this point is a saddle for 
$Q(Q+\lambda) < (3/2)(1-w_m^2)$ and a stable node 
for $Q(Q+\lambda) >(3/2)(1-w_m^2)$.
However, since $\Omega _\vp=-2Q^2/3(1-w_m)<0$ for 
$0 \le w_m<1$, 
the fixed point (a) is not physically meaningful.

\item Point (c): 

Since both $\mu_1$ and $\mu_2$ are negative 
independent of the values of $\lambda$ and $Q$, 
the point (c) is a stable node.

\item Point (d): 

{}From the condition $y^2>0$,
we require that $2Q(Q+\lambda)>3(1-w_m^2)$ for the 
existence of the critical point (d).
Under this condition we find that $\mu_1<0$
and $\mu_2>0$. 
Therefore the point (d) corresponds to a saddle.

\end{itemize}

The properties of the critical points are summarized in Table III.
The scaling solution becomes always unstable for phantom.
Therefore one can not construct a coupled dark energy 
scenario in which the present value of $\Omega_\vp$ ($\simeq 0.7$)
is a late-time attractor. 
This property is different from the case of an ordinary field
in which scaling solutions can be stable fixed points.
The only viable stable attractor for phantom is 
the fixed point (c), giving the dark energy dominated universe 
($\Omega_\vp=1$) with an equation of state 
$w_\vp=-1-\lambda^2/3<-1$.

In Figs.~\ref{phan1} and \ref{phan2} we plot the phase plane for 
two different cases. Figure \ref{phan1} corresponds to  
$Q=3/4$, $\lambda=1$, $w_m=0$ and $c=1$. In this case
the fixed point (a) is saddle, whereas the point (d) does not exist.
It is clear from Fig.~\ref{phan1} that the fixed point (c) is a global 
attractor. We also note that the allowed range in the phase plane
corresponds to $-1 \le x^2-y^2 \le 0$, which comes from the 
condition $0 \le \Omega_\vp \le 1$.
The saddle point (a) exists outside of this region.
Figure \ref{phan2} corresponds to $Q=3/2$, $\lambda=2$, $w_m=0$
and $c=1$. 
In this case the fixed point (a) is a stable node, whereas there 
exists a saddle point (d).
In Fig.~\ref{phan2} we find that the fixed points (a) and (d) are 
actually stable nodes, although the point (a) is not physically meaningful.
Compared to Fig.~\ref{phan1}, the critical point (d) newly appears,
but this is not a late-time attractor.
It is worth mentioning that numerical results agree very well 
with our analytic estimation for the stability analysis
about critical points.

\section{Dilatonic ghost condensate}

It was pointed out in Ref.~\cite{UV} that at the quantum 
level a phantom field is plagued 
by a vacuum instability associated with the production of 
ghosts and photon pairs.
We recall that the phantom field is characterised by 
a negative value of the quantity: $p_{X} \equiv \partial p/\partial X$.
One can consider a scenario which avoids the problem
of quantum instability by adding higher-order derivative 
terms that stabilize the vacuum 
so that $p_X$ becomes positive \cite{Arkani}.
In the context of low-energy effective string theory 
one may consider the following Lagrangian 
that involves a dilatonic higher-order term \cite{scorre}:
\bea
\label{gcon} 
p=-X+c e^{\lambda \vp} X^2\,,
\eea
where $c$ and $\lambda$ are both positive.
The application of this scenario to dark energy was done 
in Ref.~\cite{PT}, but the stability of critical points 
was not studied.

We obtain the pressure density $(\ref{gcon})$
by choosing the function 
\bea 
g(Y)=-1+c Y\,,
\eea
in Eq.~(\ref{lag}).
Then Eqs.~(\ref{dx}) and (\ref{dy}) yield
\bea 
\label{dxgho}
\frac{\d x}{\d N} &=&   
\frac32 x \left[1+w_m+(1-w_m) x^2(-1+cY)
-2c w_m x^2Y \right] \nonumber \\
& & +\frac{1}{1-6cY} \left[3(-1+2cY)x+
\frac{3\sqrt{6}}{2}\lambda c x^2 Y+
\frac{\sqrt{6}Q}{2} \left\{1+x^2(1-3cY)\right\} 
\right]\,, \\
\label{dygho}
\frac{\d y}{\d N} &=&   
-\frac{\sqrt{6}}{2}\lambda xy+\frac32 y
\left[1+w_m+(1-w_m)x^2
(-1+cY)-2cw_m x^2 Y\right]\,.
\eea 
By Eqs.~(\ref{Ome}) and (\ref{weff}) we find
\bea 
\label{ghose}
\Omega_\vp=x^2(-1+3c Y)\,,~~~
w_\vp=\frac{-1+c Y}{-1+3cY}\,,~~~
w_{\rm eff}=w_m-(1-w_m)x^2+(1-3w_m)c x^2Y\,.
\eea 
The stability of quantum fluctuations is ensured for
$p_X+2Xp_{XX}>0$ and $p_X \ge 0$ \cite{PT}, which
correspond to the condition $cY \ge 1/2$.
In this case one has $w_\vp \ge -1$ by Eq.~(\ref{ghose}),
which means that the presence of the term 
$e^{\lambda \vp} X^2$ leads to the stability of vacuum
at the quantum level.

\subsection{Fixed points}

Setting $\d X/\d N=0$ and $\d Y/\d N=0$ in 
Eqs.~(\ref{dxgho}) and (\ref{dygho}), we obtain 
\bea
\label{gcon2} 
y^2=-\frac{3cx^3[4-\sqrt{6}(Q+\lambda)x]}
{\sqrt{6}x (\lambda x -\sqrt{6})+\sqrt{6}Q(1+x^2)}\,,~~~~
y^2=\frac{3c(3w_m-1)x^4}{3(1+w_m)-3(1-w_m)x^2
-\sqrt{6} \lambda x}\,.
\eea
Combining these relations, we get four critical points 
presented in Table \ref{critghost}.
The point (a) corresponds to $x=0$ and $y=0$, in which 
case one has $Y \to \infty$ by Eq.~(\ref{gcon2}).
Since we require the condition $w_m \le -1$
in order to satisfy $0 \le \Omega_\vp \le 1$,
this is not a physically meaningful solution.

The points (b) and (c) correspond to 
the dark-energy dominated universe with $\Omega_\vp=1$.
Therefore these points
exist irrespective of the presence of 
the coupling $Q$.
In Table \ref{critghost} the functions $f_{\pm} (\lambda)$ 
are defined by 
\bea
\label{func}
f_{\pm} (\lambda) \equiv 1 \pm \sqrt{1+16/(3\lambda^2)}\,.
\eea
Since one has $w_{\rm eff}=w_\vp=(-1+cY)/(-1+3cY)$ 
for the points (b) and (c), 
the condition for an accelerated expansion, $w_{\rm eff}<-1/3$,
gives $cY<2/3$.
We note that $w_{\rm eff}<-1$ for $cY<1/2$ and 
$w_{\rm eff}>-1$ otherwise.
The former case corresponds to the phantom equation of state, 
whereas the latter belongs to the case in which the system is stable 
at the quantum level.
The parameter range of $Y$ for the point (b) is
$1/3<cY<1/2$, which means that 
the field $\vp$ behaves as a phantom.
In this case the universe exhibits an acceleration for any 
value of $\lambda$ and $Q$.
The point (c) belongs to the parameter range given by 
$1/2<cY<\infty$.
The condition for an accelerated expansion corresponds to
$\lambda^2 f_+(\lambda)<8/3$, which gives 
$\lambda \lesssim 0.817$.
In the limit $\lambda \to 0$ we have $cY \to 1/2$, $\Omega_\vp \to 1$
and $w_{\rm eff}=w_\vp \to -1$ for both points (b) and (c).
The $\lambda=0$ case is the original ghost condensate 
scenario proposed in Ref.~\cite{Arkani}, i.e., $p=-X+X^2$,
in which case one has an equation of 
state of cosmological constant ($w_{\vp}=-1$).

The point (d) corresponds to a scaling solution.
Since the effective equation of state is given by 
$w_{\rm eff}=(w_m \lambda- Q)/(\lambda+Q)$, 
the universe accelerates for $Q>\lambda(1+3w_m)/2$.
The requirement of the conditions $0 \le \Omega_\vp \le 1$ and
$Y=x^2/y^2>0$ places constraints on the values of $Q$ and $\lambda$.
For the non-relativistic dark matter ($w_m=0$) with 
positive coupling ($Q>0$), we have the following constraint
\bea
\label{const}
\frac12 \left[\sqrt{9Q^2+12}-5Q\right]<\lambda<\frac{1}{Q}-Q\,.
\eea
This implies that $Q$ needs to be smaller than 1
for positive $\lambda$.

\begin{table*}[t]
\begin{center}
\begin{tabular}{|c|c|c|c|c|c|}
Name & $x$ & $cY$ & $\Omega_\vp$ & $w_\vp$  
& $w_{\rm eff}$  \\
\hline
\hline
(a) & 0 & $\infty$ & $\frac{3(w_m+1)}{3w_m-1}$ &  1/3   
& $-1$ \\
\hline
(b) & $-\frac{\sqrt{6}\lambda f_+(\lambda)}{4}$ & 
$\frac12+\frac{\lambda^2 f_- (\lambda)}{16}$
& 1 & $\frac{-8+\lambda^2 f_-(\lambda)}
{8+3 \lambda^2 f_- (\lambda)}$ &
$\frac{-8+\lambda^2 f_-(\lambda)}
{8+3 \lambda^2 f_- (\lambda)}$ \\
\hline
(c) & $-\frac{\sqrt{6}\lambda f_-(\lambda)}{4}$ & 
$\frac12+\frac{\lambda^2 f_+(\lambda)}{16}$
& 1 & $\frac{-8+\lambda^2 f_+(\lambda)}{8+3 \lambda^2 f_+ (\lambda)}$ 
& $\frac{-8+\lambda^2 f_+(\lambda)}{8+3 \lambda^2 f_+ (\lambda)}$ 
\\
\hline
(d) & $\frac{\sqrt{6}(1+w_m)}{2(\lambda+Q)}$ & 
$\frac{3(1-w_m^2)-2Q(\lambda+Q)}{3(1-3w_m)(1+w_m)}$
& $\frac{3(1+w_m)[1+w_m-Q(\lambda+Q)]}{(\lambda+Q)^2(1-3w_m)}$ 
& $\frac{3(1+w_m)w_m-Q(\lambda+Q)}{3(1+w_m)-3Q(\lambda+Q)}$ 
& $\frac{w_m \lambda -Q}{\lambda +Q}$
\\
\hline
\end{tabular}
\end{center}
\caption[critghost]{\label{critghost} The critical points for the ghost condensate 
model (\ref{gcon}). Here the functions $f_{\pm} (\lambda)$ are defined 
by Eq.~(\ref{func}).
}
\end{table*}

%
\subsection{Stability around fixed points}

The elements of the matrix ${\cal M}$ for perturbations 
$u$ and $v$ are given by 
\bea 
a_{11} &=& \frac32 \left[1+w_m-3(1-w_m)x_c^2+5c(1-3w_m)x_c^2Y_c
\right]+\frac{-3+\sqrt{6}x_cQ+18cY_c+
6\sqrt{6}c(\lambda-Q)x_cY_c }{1-6cY_c} \nonumber \\
& &+\frac{12c Y_c}{x_c(-1+6cY_c)^2} 
\left[3(-1+2c Y_c)x_c+\frac{3\sqrt{6}}{2}\lambda cx_c^2Y_c
+\frac{\sqrt{6}Q}{2}
\left\{1-x_c^2(-1+3cY_c)\right\}\right]\,, \\
a_{12} &=& -3c(1-3w_m)x_c^2Y_c^{3/2}-
\frac{\sqrt{6}c}{1-6cY_c}Y_c^{3/2}
\left\{2\sqrt{6}+3x_c(\lambda-Q)\right\} \nonumber \\
& & -\frac{12cY_c}{y_c(-1+6cY_c)^2}
\left[3(-1+2cY_c)x_c+\frac{3\sqrt{6}}{2}\lambda 
c x_c^2Y_c+\frac{\sqrt{6}Q}{2}
\left\{1-x_c^2(-1+3cY_c)\right\}\right]\,, \\
a_{21} &=& 3y_c \left[-(1-w_m)x_c+
2c(1-3w_m)x_cY_c\right]
-\frac{\sqrt{6}}{2}\lambda y_c\,, \\
a_{22} &=& \frac32 \left[1+w_m-(1-w_m)x_c^2
-c(1-3w_m)x_c^2Y_c
\right]-\frac{\sqrt{6}}{2}\lambda x_c\,.
\eea 
Since the expression of the matrix elements is rather complicated, 
the eigenvalues of ${\cal M}$ are not simply written unlike 
the case of Sec.\,III.
However we can numerically evaluate $\mu_1$ and $\mu_2$ 
and investigate the stability of fixed points.

\begin{itemize}

\item Point (a): 

In this case the component $a_{12}$ diverges, which 
means that this point is unstable in addition to the fact 
$\Omega_\vp$ does not belong to the range 
$0 \le \Omega_\vp \le 1$ for plausible values of $w_m$.

\item Point (b): 

We numerically evaluate the eigenvalues $\mu_1$ and $\mu_2$
for $w_m=0$, $c=1$
and find that the point (b) is either a stable spiral or 
a stable node. When $Q \lesssim 10$ the determinant of the 
matrix ${\cal M}$ is negative with negative real parts
of $\mu_1$ and $\mu_2$, which means that the point (b) 
is a stable spiral.
In the case of $Q \gtrsim 10$, the determinant 
is positive with negative $\mu_1$ and $\mu_2$ 
if $\lambda$ is smaller than a value $\lambda_*(Q)$, 
thereby corresponding to a stable node.
Here $\lambda_*(Q)$ depends on the value $Q$.
When $\lambda>\lambda_*(Q)$ and $Q \gtrsim 10$, 
the point (b) is a stable spiral.

\item Points (c) \& (d): 

It would be convenient to discuss the stability
of these critical points together as there exists an interesting relation
between them.
We shall consider the case of $w_m=0$ and $c=1$.

For the point (c) we have numerically found that $\mu_2$ is 
always negative irrespective of the values of $Q$ and $\lambda$.
Our analysis shows that there exists a critical value $\bar{\lambda}_* (Q)$ such that
$\mu_1$ is negative for $\lambda<\bar{\lambda}_* (Q)$
and becomes positive for $\lambda>\bar{\lambda}_* (Q)$.
The critical value of  $\lambda$ can be computed
by demanding $(a_{11}a_{22}-a_{12}a_{21})=0$, which leads to
\be
\bar{\lambda}_* (Q)=\frac12 \left[\sqrt{9Q^2+12}-5Q\right]\,.
\label{lambdacrdil}
\ee
We conclude that in the region specified by $0<\lambda<\bar{\lambda}_* (Q)$, the critical
point (c) is a stable node which becomes a saddle as we move out of 
this region [$\lambda>\bar{\lambda}_* (Q)$].

For the point (d) the second eigenvalue $\mu_2$ is negative or
${\rm Re}\,(\mu_2)<0$ for all values of  $Q$ and $\lambda$. 
Meanwhile the first eigenvalue exhibits an interesting behavior. 
We recall that the allowed domain for the existence of the point (d) lies 
outside the region of stability for the point (c), see Eq.~(\ref{const}). 
If we extend it to the region $\lambda < \bar{\lambda}_* (Q)$, 
we find that point (d) is a saddle in this domain ($\mu_1>0$ and $\mu_2<0$).
The critical value of ${\lambda}$, at which $\mu_1$ for the point (d) vanishes, 
exactly coincides with $\bar{\lambda}_* (Q)$ given by Eq.~(\ref{lambdacrdil}).
As we move out of the domain of stability for the point (c), 
the critical point (d) becomes a stable node as $\mu_1<0$ in this case.
We numerically find there exists a second critical value 
$\tilde{\lambda}_{**}(Q)~(>\bar{\lambda}_* (Q))$ at which
the determinant ${\cal D}$ of the system vanishes
such that $\mu_1<0$ for $ \bar{\lambda}_* (Q)<\lambda<\tilde{\lambda}_{**}(Q)$ and 
${\rm Re}\,(\mu_1)<0$ for $\lambda >\tilde{\lambda}_{**}(Q)$.
To have an idea of orders of magnitudes, let us quote some numerical values of $ \bar{\lambda}_* (Q)$ and
$\tilde{\lambda}_{**}(Q)$, for instance $\tilde{\lambda}_{**}(Q)=2.06,~1.15$; 
$\bar{\lambda}_* (Q)=1.26,~0.64$ for $Q=0.2, 0.5$ respectively. 
The stability of (c) \& (d) can be briefly summarised as follows. 
The point (c) is a stable node
whereas the point (d) is a saddle  
for $0<\lambda<\bar{\lambda}_* (Q)$. 
The point (c) becomes a saddle for $\lambda>\lambda_* (Q)$.
In the region characterised by $\bar{\lambda}_* (Q)<\lambda<\tilde{\lambda}_{**}(Q)$, 
the critical point (d) is a stable
node but a stable spiral for $\lambda>\tilde{\lambda}_{**}(Q)$.

\end{itemize}

We summarize the property of fixed points in 
Table \ref{staghost}.
Although the point (b) is always stable at the classical level, 
this corresponds to a phantom 
equation of state ($w_\vp<-1$).
Therefore this is plagued by the instability of vacuum 
at the quantum level, 
whereas the point (c) is free from such a quantum instability. 
The scaling solution (d) also gives rise to an equation of 
state $w_\vp>-1$ as can be checked  
by the expression of $w_\vp$ in Table~\ref{critghost}.
When the point (c) is stable we find that the point (d)
is unstable, and vice versa.
Therefore the final viable attractor is described 
by a dark energy dominant universe with $\Omega_\vp=1$
[case (c)] or by a scaling solution with 
$0<\Omega_\vp={\rm const}<1$ [case (d)].

\begin{table*}[t]
\begin{center}
\begin{tabular}{|c|c|c|c|}
Name & Stability & Acceleration  & Existence \\
\hline
\hline
(a) & Unstable & All values & $w_m \le -1$ \\
\hline
(b) & Stable spiral or stable node & All values & All values  \\
\hline
(c) & Stable node or saddle & $\lambda<0.817$ & All values  \\
\hline
(d) & Stable node or stable spiral or saddle & 
$Q>\lambda (1+3w_m)/2$ & 
$[(9Q^2+12)^{1/2}-5Q]/2<\lambda<1/Q-Q$ for $w_m=0$ 
\\
\hline
\end{tabular}
\end{center}
\caption[staghost]{\label{staghost} 
The conditions for stability \& acceleration
\& existence for the ghost condensate model (\ref{gcon}). }
\end{table*}

\begin{figure}
\includegraphics[height=3.0in,width=3.2in]{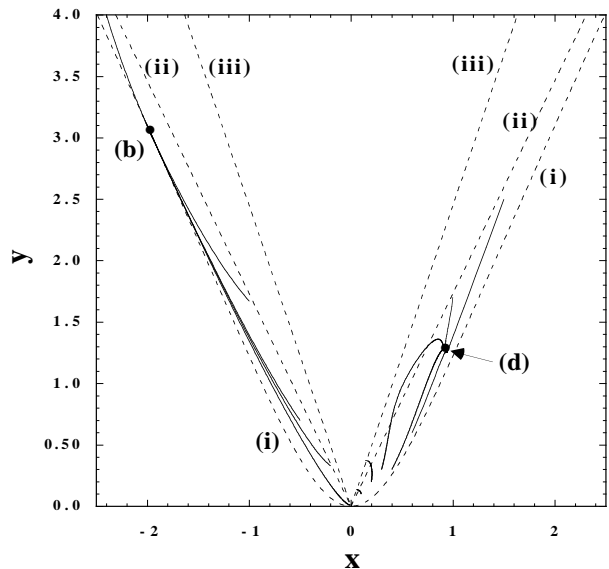}
\caption{\label{phaseghost}
Phase plane analysis in the dilatonic ghost condensate case for $Q=0.5$ and 
$\lambda=0.8$ with $w_m=0$ and $c=1$. 
In this case there exist two stable points 
(b) $(x, y)=(-1.98, 3.07)$ and (d) $(x, y)=(0.94, 1.25)$.
The curves (i), (ii) and (iii) correspond to 
$y^2=3cx^4/(1+x^2)$, $y^2=3cx^2$ and $y^2=6cx^2$, respectively.
}
\end{figure}

In Fig.~\ref{phaseghost} we plot the phase plane for 
$Q=0.5$ and $\lambda=0.4$.
In this case the point (c) is a saddle, whereas the point 
(d) is a stable node. By using the condition 
$0 \le \Omega_\vp \le 1$ in Eq.~(\ref{ghose}), we find that 
$x$ and $y$ are constrained to be in the range:
$3cx^4/(1+x^2) \le y^2  \le 3cx^2$. 
We also obtain the condition, $y^2 \le 2cx^2$,
if the stability of quantum fluctuations is taken into account \cite{PT}.
When $x$ and $y$ are initially smaller than of order unity
with positive $x$, the trajectories tend to approach 
the line $y=\sqrt{6c}x$
on which the speed of sound, $c_s \equiv \sqrt{p_X/\rho_X}$,
diverges \cite{PT}. Therefore these cases are physically 
unrealistic. Meanwhile when initial conditions of $x$ and $y$
are not much smaller than of order unity, the solutions approach
the stable point (d) provided that $x$
is positive (see Fig.~\ref{phaseghost}). 
When $x$ and $y$ are much smaller than 1 during matter
dominant era, it is difficult to reach the critical point (d)
for constant $Q$.
If the coupling $Q$ rapidly grows during the transition 
to scalar field dominated era, it is possible to approach the 
scaling solution (d) \cite{PT}.

If the initial value of $x$ is negative, the trajectories 
approach the stable point (b).
In this case we numerically found that the solutions
do not cross the lines $y=\sqrt{6c}|x|$ 
even for the initial values of $|x|$ much smaller than unity.
The final attractor point (b) corresponds to the phantom-dominant 
universe with $\Omega_\vp=1$. 

\section{Tachyon and phantom tachyon}

The Lagrangian of a tachyon field is given, in general, 
by \cite{Sen} 
\begin{eqnarray}
\label{tach}
p=-V(\phi) \sqrt{1-\epsilon \dot{\phi}^2}\,,
\end{eqnarray}
where $\epsilon=+1$.
Here we allowed the possibility of a 
phantom tachyon ($\epsilon=-1$) \cite{Hao}.
We note that there are many works in which the dynamics 
of tachyon is investigated in a cosmological 
context \cite{tachyon}.
The potential corresponding to scaling solutions
is the inverse power-law type, i.e., 
$V(\phi) \propto \phi^{-2}$.
If we choose the function $g(Y)$ as 
\begin{eqnarray}
g(Y) &=&-c\sqrt{1-2\epsilon Y}/Y\,,
\end{eqnarray}
the Lagrangian (\ref{lag}) yields $p=-ce^{-\lambda \vp}\sqrt{1-2\epsilon Y}$.
With a suitable field redefinition given by $\phi=(2/\lambda)e^{{\lambda \vp}/2}$,
we obtain the tachyon Lagrangian (\ref{tach}) with an inverse square 
potential: $V(\phi)=4c/(\lambda^2 \phi^2)$.
Note that we are considering positive $c$.

We shall introduce a new variable $\tx$ which is defined by 
$\tx^2 \equiv \dot{\phi}^2/2=Y$.
Then $x$ and $\tilde{x}$ are now related as $x=\tilde{x}y$. 
Noting that
\begin{eqnarray}
g(Y)+5Yg'(Y)+2Y^2g''(Y)&=&\frac{c\epsilon}{(1-2\epsilon Y)^{3/2}}\,,
\end{eqnarray}
we obtain the following equations by using Eqs.~(\ref{dx})
and (\ref{dy}):
\begin{eqnarray}
\label{eqtx2}
\frac{\d \tilde{x}}{\d N} &=& 
-(1-2\epsilon \tilde{x}^2) \left[ 3\tilde{x}-\frac{\sqrt{6}\lambda y}
{2\epsilon}+\frac{\sqrt{6}Q}{2c\epsilon y}
(\sqrt{1-2\epsilon \tilde{x}^2}-cy^2)\right]\,, \\
\frac{\d y}{\d N} &=& 
-\frac{\sqrt{6}}{2}\lambda \tilde{x} y^2+
\frac{3y}{2}\left[\gamma -\frac{cy^2}{\sqrt{1-2\epsilon \tilde{x}^2}}
(\gamma-2\epsilon \tilde{x}^2)\right]\,,
\end{eqnarray}
where
\begin{eqnarray}
\gamma \equiv 1+w_m\,.
\end{eqnarray}
Eqs.~(\ref{Ome}) and (\ref{weff}) give
\begin{eqnarray}
\Omega_{\vp}=\frac{c y^2}{\sqrt{1-2\epsilon \tx^2}}\,,~~~
w_{\vp}=2\epsilon \tilde{x}^2-1\,,~~~
w_{\rm eff}=w_m-\frac{cy^2(1+w_m-2\epsilon \tx^2)}
{\sqrt{1-2\epsilon \tx^2}}\,.
\end{eqnarray}

\subsection{Fixed points}

When we discuss fixed points in the tachyon model, it may be convenient 
to distinguish between two cases: $Q=0$ and $Q \ne 0$.
We note that the fixed points were derived in Ref.~\cite{AL,CGST}
for $Q=0$ and $\epsilon=+1$.

\subsubsection{$Q = 0$}

We summarize the fixed points in Table \ref{crittach1} for $Q=0$.
The point (a) is a fluid-dominant solution ($\Omega_m \to 1$)
with an effective equation of state, $w_{\rm eff}=w_m$.
Then the accelerated expansion does not occur unless
$w_m$ is less than $-1/3$.
The points (b1) and (b2) are the kinematic dominant solution whose 
effective equation of state corresponds to a dust.
This does not exist in the case of phantom tachyon.
The point (c) is a scalar-field dominant solution that gives 
an accelerated expansion at late times for 
$\lambda^2 y_c^2/(3\epsilon)<2/3$, where $y_c$ is defined by 
\begin{eqnarray}
\label{yc}
y_c \equiv \left[ \frac{-\lambda^2 \pm \sqrt{\lambda^4+36c^2}}
{6\epsilon c^2} \right]^{1/2}\,.
\end{eqnarray}
This condition translates into $\lambda^2<2\sqrt{3}c$ 
for $\epsilon=+1$.
In Eq.~(\ref{yc}) the plus sign corresponds to $\epsilon>0$, whereas 
the minus sign to $\epsilon<0$.
Note that phantom tachyon gives an effective equation of 
state $w_{\rm eff}$ that is smaller than $-1$.
The points (d1) and (d2) correspond to scaling solutions in which 
the energy density of the scalar field decreases proportionally to 
that of the perfect fluid ($w_\vp=w_m$).
The existence of this solution requires the condition $w_m<0$
as can be seen in the expression of $\Omega_\vp$.
We note that the fixed points (d1) and (d2) do not 
exist for phantom tachyon unless the background fluid
behaves as phantom ($w_m<-1$).

\begin{table*}[t]
\begin{center}
\begin{tabular}{|c|c|c|c|c|c|}
Name & $\tx$ & $y$ & $\Omega_\vp$ & $w_\vp$ & 
$w_{\rm eff}$  \\
\hline
\hline
(a) & 0 & 0 & 0 & $-1$ & $w_m$ \\
\hline
(b1) & $\frac{1}{\sqrt{2\epsilon}}$ & 0 & 1 & 0 &
0 \\
\hline
(b2) & $-\frac{1}{\sqrt{2\epsilon}}$ & 0 & 1 & 0 & 
0 \\
\hline
(c) & $\frac{\lambda y_c}{\sqrt{6}\epsilon}$ & $y_c$
 & 1 & $\frac{\lambda^2 y_c^2}{3\epsilon}-1$ &
$\frac{\lambda^2 y_c^2}{3\epsilon}-1$ \\
\hline
(d1) & $(\gamma/2\epsilon)^{1/2}$ & 
$\frac{\sqrt{3\epsilon \gamma}}{\lambda}$ & 
$\frac{3c\epsilon \gamma}{\lambda^2 \sqrt{1-\gamma}}$ & 
$w_m$ & $w_m$ \\
\hline
(d2) & $-(\gamma/2\epsilon)^{1/2}$ & 
$-\frac{\sqrt{3\epsilon \gamma}}{\lambda}$ & 
$\frac{3c\epsilon \gamma}{\lambda^2 
\sqrt{1-\gamma}}$ & 
$w_m$ & $w_m$ \\
\hline
\end{tabular}
\end{center}
\caption[crittach1]{\label{crittach1} 
The critical points for the tachyon model (\ref{tach})
with $Q=0$.
Here $y_c$ is defined by Eq.~(\ref{yc}).
The points (b) and (d) do not exist for the phantom 
tachyon ($\epsilon<0$).
}
\end{table*}

\subsubsection{$Q \ne 0$}

Let us next discuss the case with $Q \ne 0$.
The critical points are summarized in Table~\ref{crittach2}.
We first note that the fixed point $(x, y)=(0, 0)$ disappears
in the presence of the coupling $Q$.
The points (b) and (c) appear as is similar to the case $Q=0$.
While (b1) and (b2) do not exist for phantom tachyon, 
the point (c) exists both for $\epsilon>0$ and $\epsilon<0$.

The point (d) [$(x_i, y_i)$] corresponds to scaling solutions, whose 
numbers of solutions depend on the value of $w_m$.
$y_i$ are related with $x_i$ through the relation 
$y_i=3\gamma/[\sqrt{6}(\lambda+Q)\tx_i]$.
Here $x_i$ satisfy the following equation 
\begin{eqnarray}
\label{xi}
\frac{\gamma-2 \epsilon \tx_i^2}{\tx_i^2 
\sqrt{1-2\epsilon \tx_i^2}}=
\frac{2Q(\lambda+Q)}{3(1+w_m)c}\,.
\end{eqnarray}
We note that only positive $x_i$ are allowed
for $\lambda>0$, since $y_i$ is positive definite.
If we introduce the quantity 
$\xi \equiv \sqrt{-w_\vp}=\sqrt{1-2\epsilon \tx_i^2}$, we find 
\begin{eqnarray}
\label{function}
f(\xi) \equiv \frac{\xi^2+w_m}{\xi (1-\xi^2)}
=\frac{Q(Q+\lambda)}{3c\epsilon (1+w_m)}\,.
\end{eqnarray}
The behavior of the function $f(\xi)$ is different depending on 
the value of $w_m$.
We note that $\lambda$ is related with the slope of the 
potential as $\lambda=-V_\phi/V^{3/2}$ \cite{CGST}, 
which means that $\lambda>0$ for the inverse 
power-law potential: $V(\phi)=M^2\phi^{-2}$.
Then the r.h.s. of Eq.~(\ref{function})
is positive for $\epsilon>0$ and negative otherwise.
The allowed range of $\xi$ is $0<\xi<1$ for $\epsilon>0$
and $\xi>1$ for $\epsilon<0$.
We can classify the situation as follows.

\begin{itemize}

\item $w_m>0$: 

The function $f(\xi)$ goes to infinity for $\xi \to 0+0$ and $\xi \to 1-0$.
It has a minimum at $\xi=\xi_M$ with $0<\xi_M<1$.
Here $\xi_M$ is defined by 
$\xi_M^2 \equiv [-(1+3w_m)+\sqrt{(1+w_m)(1+9w_m)}]/2$.
Then for $\epsilon>0$, there exist two solutions for Eq.~(\ref{function})
provided that $Q(Q+\lambda)/(3c\epsilon (1+w_m))>f(\xi_M)$.
The function $f(\xi)$ has a dependence $f(\xi) \to -\infty$ for
$\xi \to 1+0$ and $f(\xi) \to 0$ for $\xi \to +\infty$.
Then for $\epsilon<0$, there exists one scaling solution for 
Eq.~(\ref{function}).

\item $w_m=0$:

In this case we can analytically derive the solution for Eq.~(\ref{function}). 
The function $f(\xi)$ is zero at $\xi=0$ and monotonically increases
toward $+\infty$ as $\xi \to 1-0$.
Then we have one solution for Eq.~(\ref{function}) if $\epsilon>0$.
The function $f(\xi)$ has a dependence $f(\xi) \to -\infty$ for
$\xi \to 1+0$ and $f(\xi) \to 0$ for $\xi \to +\infty$.
This again shows the existence of one solution for 
Eq.~(\ref{function}) if $\epsilon<0$.
The solutions for Eq.~(\ref{function}) are given by 
$\xi=[-1 + \sqrt{1+4A^2}]/2A$ for $\epsilon=+1$ and 
$\xi=[1 + \sqrt{1+4A^2}]/2A$ for $\epsilon=-1$, where
$A \equiv Q(Q+\lambda)/(3c(1+w_m))$.
Then we obtain the following fixed points
together with the equation of state $w_\vp$:
\begin{eqnarray}
\label{crixc}
& &\tilde{x}_c= \frac{\sqrt{3}\sqrt{c(-3c+\sqrt{9c^2+4Q^2(\lambda+Q)^2})}}{2Q(\lambda+Q)} \,, \\
\label{criyc}
& &y_c= \frac{\sqrt{6}Q}{\sqrt{3}\sqrt{c(-3 c+\sqrt{9c^2+4Q^2(\lambda+Q)^2}})} \,, \\
& & w_\vp=-\frac{9c^2}{4Q^2(Q+\lambda)^2}
\left[\sqrt{1+\frac{4Q^2(Q+\lambda)^2}{9c^2}}-1
\right]^2\,,
\end{eqnarray}
for $\epsilon=+1$ and 
\begin{eqnarray}
\label{crixc2}
& & \tx_c= \frac{\sqrt{3}\sqrt{c(3 c+\sqrt{9c^2+4Q^2(\lambda+Q)^2})}}{2Q(\lambda+Q)} \,, \\
\label{criyc2}
& & y_c= \frac{\sqrt{6}Q}{\sqrt{3}\sqrt{c(3\ c+\sqrt{9c^2+4Q^2(\lambda+Q)^2}})}\,, \\
& &w_\vp=-\frac{9c^2}{4Q^2(Q+\lambda)^2}
\left[\sqrt{1+\frac{4Q^2(Q+\lambda)^2}{9c^2}}+1
\right]^2\,,
\end{eqnarray}
for $\epsilon=-1$.

For ordinary tachyon one has $\tx_c \to 1/\sqrt{2}$, 
$y_c \to \sqrt{3}/(\lambda+Q)$, $w_\vp \to 0$
as $Q \to 0$ and $\tx_c \to 0$, 
$y_c \to 1/\sqrt{c}$, $w_\vp \to -1$ as $Q \to +\infty$.
There is a critical value $Q_*(\lambda)$ which gives 
the border of acceleration and deceleration, i.e., 
the accelerated expansion occurs for $Q>Q_*(\lambda)$.
For phantom tachyon we obtain $\tx_c \to \infty$, 
$y_c \to 0$, $w_\vp \to -\infty$
as $Q \to 0$ and $\tx_c \to 0$, 
$y_c \to 1/\sqrt{c}$, $w_\vp \to -1$ as $Q \to +\infty$.
Thus the presence of the coupling $Q$ can lead to an accelerated
expansion. 

\item $w_m<0$: 

The function $f(\xi)$ is zero at $\xi=\sqrt{-w_m}$.
It monotonically 
increases toward $+\infty$ as $\xi \to 1-0$
in the region $\sqrt{-w_m}<\xi<1$.
This means that we have one solution for Eq.~(\ref{function}) 
if $\epsilon>0$.
The function $f(\xi)$ has a dependence $f(\xi) \to -\infty$ for
$\xi \to 1+0$ and $f(\xi) \to 0$ for $\xi \to +\infty$, which shows
the existence of one solution for Eq.~(\ref{function}) 
if $\epsilon<0$.
Note, however, that the $w_m<0$ case is not realistic.

\end{itemize}

\begin{table*}[t]
\begin{center}
\begin{tabular}{|c|c|c|c|c|c|}
Name & $\tx$ & $y$ & $\Omega_\vp$ & $w_\vp$ & 
$w_{\rm eff}$  \\
\hline
\hline
(b1) & $\frac{1}{\sqrt{2\epsilon}}$ & 0 & 1 & 0 &
0 \\
\hline
(b2) & $-\frac{1}{\sqrt{2\epsilon}}$ & 0 & 1 & 0 & 
0 \\
\hline
(c) & $\frac{\lambda y_c}{\sqrt{6}\epsilon}$ & $y_c$
 & 1 & $\frac{\lambda^2 y_c^2}{3\epsilon}-1$ &
$\frac{\lambda^2 y_c^2}{3\epsilon}-1$ \\
\hline
(d) & $\tx_i$ & 
$\frac{3\gamma}{\sqrt{6}(\lambda+Q)\tx_i}$ & 
$\frac{3c\gamma^2}{2(\lambda+Q)^2 \tx_i^2 \sqrt{1-2\epsilon \tx_i^2}}$ 
& $2\epsilon \tx_i^2-1$ & $w_m-\frac{cy_i^2(1+w_m-2\epsilon \tx_i^2)}
{\sqrt{1-2\epsilon \tx_i^2}}$ \\
\hline
\end{tabular}
\end{center}
\caption[crit]{\label{crittach2} 
The critical points for the tachyon model
with $Q \ne 0$.
Here $y_c$ is defined by Eq.~(\ref{yc}).
$\tx_i$ are the solutions of Eq.~(\ref{xi}).
The critical points (b1) and (b2) do not exist 
for the phantom tachyon ($\epsilon<0$).
The numbers of the point (d) depend on the value of $w_m$. 
}
\end{table*}

\subsection{Stability}

The components of the matrix ${\cal M}$ are
\begin{eqnarray}
a_{11}&=& -3+18\,\epsilon \tx_c^{2}-2\sqrt{6}
(\lambda+Q) \tx_c y_c+
\frac{3\sqrt{6}Q\tx_c}{cy_c}\,\sqrt{1-2\epsilon \tx_c^2}\,, \\
a_{12}&=& (1-2\epsilon \tx_c^2) \left[ \frac{\sqrt{6}(\lambda+Q)}
{2\epsilon}+\frac{\sqrt{6}Q}{2c\epsilon y_c^2} 
\sqrt{1-2\epsilon \tx_c^2} \right]\,,\\
a_{21}&=&-\frac{\sqrt {6}}{2}\lambda y_c^2-
\frac{3c \epsilon  \tx_c y_c^3 (\gamma-2 \epsilon \tx_c^2)}
{(1-2\epsilon \tx_c^2)^{3/2}}+
\frac{6c\epsilon \tx_c y_c^3}{\sqrt{1-2\epsilon \tx_c^2}}\,, \\
 a_{22}&=&-\sqrt {6}\lambda\,\tx_cy_c+\frac32 \gamma-
 \frac{9cy_c^2 (\gamma- 2 \epsilon \tx_c^2)}
 {2\sqrt{1-2\epsilon \tx_c^2}}\,.
\end{eqnarray}
Hereafter we shall discuss the stability of fixed points 
for an ordinary tachyon ($\epsilon=+1$)
and for a phantom tachyon ($\epsilon=-1$)
by evaluating eigenvalues of the matrix ${\cal M}$.

\subsubsection{Ordinary tachyon $(\epsilon=+1)$}

The stability of fixed points is summarized 
in Table \ref{tachcrit2}.

\begin{itemize} 

\item Point (a): 

This point exists only for $Q=0$.
The eigenvalues are 
\begin{eqnarray}
\mu_1=3\gamma/2\,,~~~\mu_2=-3\,.
\end{eqnarray}
Therefore the point (a) is a saddle for $\gamma>0$
and a stable node for $\gamma<0$.
Therefore this point is not stable
for an ordinary fluid satisfying $\gamma \ge 1$.

\item Points (b1) and (b2): 

Since the eigenvalues are
\begin{eqnarray}
\mu_1=6\,,~~~\mu_2=9/2-3\gamma\,,
\end{eqnarray}
the points (b1) and (b2) are unstable nodes
for $\gamma<3/2$ and saddle points for $\gamma>3/2$.

\item Point (c): 

The eigenvalues are
\begin{eqnarray}
\mu_1=-3\gamma+\frac{\lambda (\lambda+Q)}{6c^2}
\left(\sqrt{\lambda^4+36c^2}-\lambda^2\right)\,,~~~~
\mu_2=-3+\frac{\lambda^2}{12c^2}
\left(\sqrt{\lambda^4+36c^2}-\lambda^2\right)\,.
\end{eqnarray}
The range of $\mu_2$ is $-3 \le \mu_2 <-3/2$.
We also find that $\mu_1 \le 0$ if
\begin{eqnarray}
\label{gammas}
\gamma \ge \gamma_s \equiv 
\frac{\lambda (\lambda+Q)}{18c^2}
\left(\sqrt{\lambda^4+36c^2}-\lambda^2\right)\,,
\end{eqnarray}
and $\mu_1>0$ if $\gamma<\gamma_s$.
Therefore the point (c) is a stable node for $\gamma \ge \gamma_s$
and a saddle point for $\gamma<\gamma_s$.
 
\item Point (d): 

When $Q=0$ we obtain 
$\gamma \le \gamma_s=\lambda^2 [\sqrt{\lambda^4+36c^2}-\lambda^2]/(18c^2)$
from the condition $\Omega_\vp \le 1$.
The eigenvalues for $Q=0$ are 
\begin{eqnarray}
\mu_{1, 2}=\frac34 \left[\gamma-2 \pm 
\sqrt{17\gamma^2-20\gamma+4+\frac{48c\gamma^2}{\lambda^2}
\sqrt{1-\gamma}}\right]\,,
\end{eqnarray}
which are both negative for $\gamma \le \gamma_s$.
Therefore the point (d) is a stable node for $Q=0$.

This situation changes if we account for the coupling $Q$.
In what follows we shall consider the case of 
a non relativistic dark matter ($w_m=0$).
The analytic expressions for the eigenvalues are rather cumbersome 
for general $Q$, but they take simple forms
in the large coupling limit ($Q \to +\infty$):
\bea
\mu _{1, 2} \simeq -3 \pm \sqrt{-\frac{3\lambda Q}{c}}\,.
\eea
This demonstrates that the critical point (d) is a stable spiral 
for large $Q$.
In fact we numerically confirmed that the determinant of the 
matrix ${\cal M}$ changes from positive to negative 
when $Q$ becomes larger 
than a critical value $Q_1(\lambda)$.
This critical value depends on $\lambda$, e.g., $Q_1(\lambda=0.1)=37.5$ 
and $Q_1(\lambda=1.0)=3.28$ for $c=1$. 
When $Q>Q_1(\lambda)$ numerical calculations show that 
the point (d) is a stable spiral for any $\lambda$.

When $Q<Q_1(\lambda)$ we find that both 
$\mu_1$ and $\mu_2$ are negative when $Q$ is larger than 
a critical value $Q_2(\lambda)$.
Meanwhile $\mu_1>0$ and $\mu_2<0$ for $Q<Q_2(\lambda)$.
Here $Q_2(\lambda)$ can be analytically derived as
\begin{eqnarray}
Q_2(\lambda)=  -\frac{\lambda}{2}+\frac{\sqrt{\lambda^4+36c^2}}{2\lambda} \,,
\end{eqnarray}
which corresponds to the eigenvalue: $\mu_1=0$.
For example we have $Q_2(\lambda=0.1)=29.95$ 
and $Q_2(\lambda=1.0)=2.54$ for $c=1$. 
{}From the above argument the fixed point (d) is a saddle 
for $Q<Q_2(\lambda)$, a stable node 
for $Q_2(\lambda)<Q<Q_1(\lambda)$ and 
a stable spiral for $Q>Q_1(\lambda)$.

\end{itemize}

\begin{table*}[t]
\begin{center}
\begin{tabular}{|c|c|c|c|}
Name & Stability & Acceleration  & Existence \\
\hline
\hline
(a) & Saddle point for $\gamma>0$  
& $\gamma<2/3$ & $Q \ne 0$
\\
&  Stable node for $\gamma<0$ & &
\\
\hline
(b1), (b2) & Unstable node for $\gamma<3/2$ & 
No & All values
\\
&  Saddle point for $\gamma>3/2$ & &
\\
\hline
(c) & Stable node for $\gamma \ge \gamma_s$ & 
$\lambda^2<2\sqrt{3}c$ & 
All values
\\
&  Saddle point for $\gamma<\gamma_s$ & &
\\
\hline
(d) [$Q=0$] & Stable node or stable spiral 
 & $\gamma<2/3$ & $0 \le \gamma<1$ 
\\
\hline
(d) [$Q \ne 0$] & Stable node or stable spiral or saddle 
 & $Q>\tilde{Q}_*(\lambda)$ & 
$\tilde{x}_i^2(1-2\tilde{x}_i^2)^{1/2} 
\ge \frac{9c\lambda^2}
{2(\lambda+Q)^2}$
\\
\hline
\end{tabular}
\end{center}
\caption[tachcrit2]{\label{tachcrit2}
The conditions for stability \& acceleration \& existence  
for an ordinary tachyon field ($\epsilon=+1$).
$\gamma_s$ is defined by Eq.~(\ref{gammas}).
$\tilde{Q}_*(\lambda)$ depends on $\lambda$.
}
\end{table*}

In the case of $w_m=0$, the stability condition 
(\ref{gammas}) for the point (c) corresponds to 
\bea
Q \le  -\frac{\lambda}{2}+\frac{\sqrt{\lambda^4+36c^2}}{2\lambda}\,.
\eea
The r.h.s. completely coincides with $Q_2(\lambda)$.
This means that the critical point (c) presents a stable node in
the region where (d) is a saddle.
When $Q>Q_2(\lambda)$ the point (d) is stable, whereas
(c) is a saddle.
Therefore one can not realize the situation in which both (c) 
and (d) are stable.

\begin{figure}
\includegraphics[height=3.0in,width=3.2in]{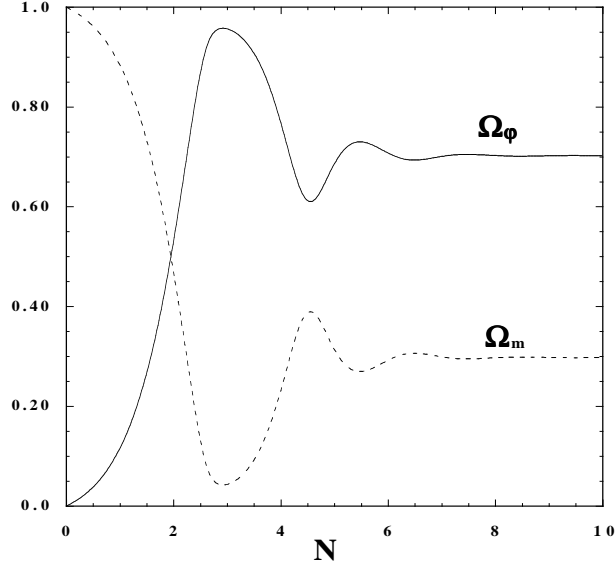}
\caption{\label{Omephim}
The evolution of $\Omega_\vp$ and $\Omega_m$ for an 
ordinary tachyon ($\epsilon=+1$) with $Q=3.0$ and $\lambda=2.18$.
The stable attractor in this case is the fixed point (d), giving  
$\Omega_\vp=0.7$ and $\Omega_m=0.3$ asymptotically.
}
\end{figure}

In Fig.~\ref{Omephim} we plot the evolution of $\Omega_\vp$ and $\Omega_m$ 
for $Q=3.0$ and $\lambda=2.18$. Since $Q_1(\lambda)=1.02$ and 
$Q_2(\lambda)=0.67$ in this case, the fixed point (d) is a stable spiral whereas
the point (c) is a saddle. In fact the solutions approach the point (d) with 
oscillations as is clearly seen in Fig.~\ref{Omephim}. 
The attractor corresponds to a scaling solution 
that gives $\Omega_\vp=0.7$ and $\Omega_m=0.3$.

%
\subsubsection{Phantom tachyon $(\epsilon=-1)$}

For a phantom tachyon the stability of fixed points exhibits
a number of differences compared to an ordinary tachyon.

\begin{itemize} 

\item Point (a): 

The property of this fixed point is completely the same as 
in the case of an ordinary tachyon.

\item Point (c): 

In this case the calculation of the eigenvalues of the matrix ${\cal M}$
is more involved compared to the point (c) for $\epsilon=+1$, but we
numerically find that $\mu_1$ and $\mu_2$ are both negative
for any value of $Q$ and $\lambda$.
Then the point (c) is a stable node.

\item Point (d): 

In the case of $w_m=0$, the critical points are given by
Eqs.~(\ref{crixc2}) and (\ref{criyc2}).
In the large coupling limit $Q \to \infty$, 
the eigenvalues are approximately given by 
\bea
\mu _{1, 2} \simeq -3 \pm \sqrt{\frac{3\lambda Q}{c}}\,,
\eea
which means that $\mu_1$ is positive and $\mu_2$ is negative.
Numerically we find that $\mu_1>0$ and $\mu_2<0$
for any value of $Q$ and $\lambda$.
Then the fixed point (d) is always saddle.

\end{itemize} 

{}From the above argument scaling solutions do not give
a viable late-time attractor for the phantom.
This property is similar to the ordinary phantom field 
discussed in Sec.~III.
Thus the solutions approach the phantom dominant universe
($\Omega_\vp=1$) even in the presence of the coupling $Q$.

\begin{table*}[t]
\begin{center}
\begin{tabular}{|c|c|c|c|}
Name & Stability & Acceleration  & Existence \\
\hline
\hline
(a) & Saddle point for $\gamma>0$  
& $\gamma<2/3$ & $Q = 0$
\\
&  Stable node for $\gamma<0$ & &
\\
\hline
(c) & Stable node & All values & All values
\\
\hline
(d) & Saddle point & All values & 
$Q \ne 0$ and $\tilde{x}_i^2
(1+2\tilde{x}_i^2)^{1/2} \ge \frac{9c\lambda^2}
{2(\lambda+Q)^2}$
\\
\hline
\end{tabular}
\end{center}
\caption[ptach]{\label{ptach} 
The conditions for stability \& acceleration \& existence  
for a phantom tachyon field ($\epsilon=-1$).
$\tilde{x}_i$ are the solutions of Eq.~(\ref{xi}).
}
\end{table*}

\section{Summary}

In this paper we studied coupled dark energy scenarios in the 
presence of a scalar field $\vp$ coupled to a barotropic perfect fluid. 
The condition for the existence of scaling solutions restricts
the form of the field Lagrangian to be 
$p=X g(Xe^{\lambda \vp})$,
where $X=-g^{\mu\nu} \partial_\mu \vp
\partial_\nu \vp /2$, $\lambda$ is a constant and 
$g$ is an arbitrary function \cite{PT}.
This Lagrangian includes a wide variety of scalar-field 
dark energy models such as quintessence,
diatonic ghost condensate, tachyon and k-essence \cite{TS}.
Our main aim was to investigate, in detail, the properties of critical points
which play crucially important roles when we construct 
dark energy models interacting with dark matter.

We first derived differential equations (\ref{dx}) and (\ref{dy})
in an autonomous form for the general Lagrangian  
$p=X g(Xe^{\lambda \vp})$ by introducing dimensionless
quantities $x$ and $y$ defined in Eq.~(\ref{xandy}).
These equations can be used for
any type of scalar-field dark energy models which possess
cosmological scaling solutions.
We note that the quantity $\lambda$ is typically related with 
the slope of the scalar-field potential $V(\vp)$, e.g., 
$\lambda \propto -V_\vp/V$ for an ordinary field \cite{scaling} and 
$\lambda \propto -V_\vp/V^{3/2}$ for a tachyon 
field \cite{CGST}.
Scaling solutions are characterized by constant $\lambda$,
thus corresponding to an exponential potential 
$V=V_0e^{-\lambda \vp}$ for an ordinary field and 
an inverse power-law potential $V=V_0\vp^{-2}$ 
for tachyon. Even for general potentials one can perform 
a phase-space analysis by considering ``instantaneous''
critical points with a dynamically changing 
$\lambda$ \cite{scaling,CGST}.
Thus the investigation based upon constant $\lambda$
contains a fundamental structure of 
critical points in scalar-field dark energy models.

We applied our autonomous equations to several different 
dark energy models--(i) ordinary scalar field (including phantom),
(ii) dilatonic ghost condensate, and (iii) tachyon (including 
phantom). In all cases we found critical points
corresponding to a scalar-field dominant solution
($\Omega_\vp=1$) with an equation of state: 
$w_{\vp} \to -1$ as $\lambda \to 0$.
These points exist irrespective of the presence of the 
coupling $Q$. This can be understood by the fact that 
the $Q$-dependent term in Eq.~(\ref{dx}) vanishes
for $\Omega_\vp \to 1$.
In the case where $w_\vp>-1$, 
these solutions are either stable nodes or saddle points  
depending on the values of $\lambda$ and $Q$,
see (d) in Table II, (c) in Table V, and (c) in Table VIII.
We note that the condition for an accelerated expansion 
requires that $\lambda$ is smaller than a critical value
$\tilde{\lambda}$, i.e., $\lambda<\sqrt{2}$ for the ordinary 
field, $\lambda<0.817$ for the dilatonic ghost condensate,
and $\lambda^2<2\sqrt{3}c$ for the tachyon.
The current universe can approach this scalar-field
dominated fixed point with an accelerated expansion 
provided that this point is a stable node and
$\lambda$ is smaller than $\tilde{\lambda}$.  
In the case of a phantom field ($w_\vp<-1$), the 
$Q$-independent solutions
explained above are found to be always stable at the classical level,
see (d) in Table III, (b) in Table V, and (c) in Table IX.
Thus the solutions tend to approach these fixed points
irrespective of the values of $\lambda$ and $Q$.
Nevertheless we need to keep in mind that this classical stability 
may not be ensured at the quantum level because of 
the vacuum instability under the production of 
ghosts and photon pairs \cite{UV,PT}. 
 
In the presence of the coupling $Q$ there exist viable
scaling solutions that provide an accelerated expansion, 
while it is not possible for $Q=0$ unless $w_m$ is 
less than $-1/3$.
If $w_\vp>-1$, the scaling solution is a stable node
or a stable spiral for an ordinary scalar field under the condition 
$\Omega_\vp<1$, see (e) in Table II.
In the cases of dilatonic ghost condensate and tachyon, 
the scaling solution can be a stable node or a stable spiral 
or a saddle depending on the values of $\lambda$ and $Q$, 
see (d) in Table V and (d) in Table VIII.
The accelerated expansion occurs 
when the coupling $Q$ is larger than a value 
$\tilde{Q}_*(\lambda)$,
e.g., $Q>\lambda (1+3w_m)/2$ for the ordinary field and 
the dilatonic ghost condensate.
When $w_\vp>-1$ we find that the scaling solution is stable 
if the scalar-field dominated fixed point ($\Omega_\vp=1$)
is unstable, and vice versa. This property holds in all models
considered in this paper.
Therefore the final attractor is either the scaling solution 
with constant $\Omega_\vp$ satisfying $0<\Omega_\vp<1$
or the scalar-field dominant solution with $\Omega_\vp=1$.

For the ordinary phantom and the phantom tachyon
we found that scaling solutions
always correspond to saddle points, 
see (e) in Table III and (d) in Table IX.
Therefore they can not be late-time attractors
unlike the case of $w_\vp>-1$.
The situation is similar in the dilatonic ghost condensate model
as well, since the solutions do not reach a scaling solution 
for initial values of $|x|$ and $y$ much smaller than 1.
The only viable stable attractor is the scalar-field 
dominant fixed point corresponding to $\Omega_\vp=1$
and $w_\vp<-1$.
Therefore the universe finally approaches a
state dominated by the phantom field even in the presence of the 
coupling $Q$.
This tells us how phantom is strong to over dominate the universe!

Since our paper provides a general formalism applicable to a variety of 
scalar fields coupled to dark matter, we hope that it would be useful 
for the concrete model building of dark energy. 
Recently Amendola et al. \cite{Amendola04} carried out a likelihood 
analysis using the dataset of Type Ia supernovae for the model 
(\ref{normalp}) and placed constraints on the values 
of $w_{\rm eff}$ and $\Omega_{\vp}$.
They found that the coupled dark energy scenario is compatible with 
the observational data only if the equation of state satisfies 
$w_{\rm eff}>-1$.
It would be quite interesting to extend the analysis to general 
coupled dark energy models presented in this paper.

\section*{ACKNOWLEDGEMENTS}
We thank N.~Dadhich, T.~Padmanabhan and V.~Sahni for
useful discussions.
B.\,G. is supported by the Thailand Research Fund (TRF).


\end{document}